\documentclass[12pt,preprint]{aastex}
\shorttitle{The Pulsating Pulsar Magnetosphere}
\shortauthors{Tsui}

\begin{document}

\title{The Pulsating Pulsar Magnetosphere}
\author{K.H. Tsui}
\affil{Instituto de F\'{i}sica - Universidade Federal Fluminense
\\Campus da Praia Vermelha, Av. General Milton Tavares de Souza s/n
\\Gragoat\'{a}, 24.210-346, Niter\'{o}i, Rio de Janeiro, Brasil.}
\email{tsui$@$if.uff.br}
\pagestyle{myheadings}
\baselineskip 18pt
	 
\begin{abstract}

Following the basic principles
 of a charge separated pulsar magnetosphere \citep{goldreich1969},
 we consider the magnetosphere be stationary in space,
 instead of corotating,
 and the electric field be uploaded
 from the potential distribution on the pulsar surface,
 set up by the unipolar induction.
 Consequently, the plasma of the magnetosphere
 undergoes guiding center drifts of the gyro motion
 due to the transverse forces to the magnetic field.
 These forces are the electric force,
 magnetic gradient force, and field line curvature force.
 Since these plasma velocities are of drift nature,
 there is no need to introduce an emf along the field lines,
 which would contradict the
 $E_{\parallel}=\vec E\cdot\vec B=0$ plasma condition.
 Furthermore, there is also no need
 to introduce the critical field line
 separating the electron and ion open field lines.
 We present a self-consistent description
 where the magnetosphere is described
 in terms of electric and magnetic fields
 and also in terms of plasma velocities.
 The fields and velocities are then connected
 through the space charge densities self-consistently.
 We solve the pulsar equation analytically for the fields
 and construct the standard steady state pulsar magnetosphere.
 By considering the unipolar induction inside the pulsar
 and the magnetosphere outside the pulsar
 as one coupled system,
 and under the condition that the unipolar pumping rate
 exceeds the Poynting flux in the open field lines,
 plasma pressure can build up in the magnetosphere,
 in particular in the closed region.
 This could cause a periodic openning up of the closed region,
 leading to a pulsating magnetosphere,
 which could be an alternative for pulsar beacons.
 The closed region can also be openned periodically
 by the build-up of toroidal magnetic field
 through a positive feedback cycle.
 
\end{abstract}
\keywords{(stars:) pulsars : general}

\maketitle

\newpage
\section{Rotating Magnetosphere}

The basic model of a pulsar had long been recognized
 as a perfect conductor (neutron star)
 rotating with an angular velocity $\vec\Omega_{0}$
 in a hiper magnetic field $\vec B$
 \citep{gold1968}.
 Considering the rotational axis and the magnetic axis
 both aligned in the same sense,
 charged particles in the neutron star
 are driven by the magnetic force to the surface
 according to their signs,
 with electrons towards the polar regions
 and ions towrds the equatorial region.
 This is the Faraday unipolar induction
 that operates homopolar generator,
 which extracts the energy of a conducting rigid rotator.
 Due to its rotation in the presence of a magnetic field,
 the conducting pulsar surface
 is not an equipotential surface,
 contrary to the static case.
 If the pulsar were surrounded by a vacuum,
 the separated charges would accumulate
 on the pulsar surface by their signs,
 and a counter electric field
 would build up inside the neutron star
 to counteract the magnetic force.
 A steady state surface charge distribution
 or potential distribution would be reached
 when the two forces inside the pulsar cancel out,
 and there would be a quadrupole electric field
 in space surrounding the neutron star.
 The unipolar induction would cease to operate then.
 However, the seminal work of \citet{goldreich1969}
 noted that such a vacuum configuration was unstable,
 and there had to be a pulsar magnetosphere
 to channel the surface charges.
 As a result, the counter electric field
 inside the pulsar would be small,
 and the unipolar induction would continue to operate.
 The steady state potential distribution on the pulsar surface
 would be uploaded to the equipotential magnetic field lines
 of the magnetosphere.
 Quite often, field line breaking near the pulsar
 is invoked to provide an electron-positron
 background pair plasma to fill the magnetosphere.

The equation of motion of a charge particle
 in the magnetosphere is given by
 
\begin{eqnarray}
\nonumber
\rho{d\vec v\over dt}\,
 =\,\rho_{q}\vec E + \vec J\times\vec B
 -\nabla p
 -\nu\rho\vec v +\rho\vec g\,\,\,.
\end{eqnarray}
  
\noindent With the plasma inertia on the left side
 and plasma pressure, collision,
 gravitational force on the right side neglected,
 comparing to the hiper magnetic and electric fields,
 the macroscopic structure of the magnetosphere
 is determined only by the balance
 between electric and magnetic forces,
 which is the force-free field model
 \citep{scharlemann1973, okamoto1974}
 with

\begin{eqnarray}
\label{eqno1}
\rho_{q}\vec E + \vec J\times\vec B\,
 =\,0\,\,\,.
\end{eqnarray}

\noindent Although plasma inertia and plasma pressure
 are neglected in the macroscopic description,
 these terms in the equation of motion
 give rise charged particle gyro motion
 about magnetic field lines
 and transverse drift motions across field lines
 in the microscopic single particle description
 \citep{thompson1962, schmidt1966, jackson1975}.
 As a result of these drifts,
 macroscopic drift currents and plasma flows,
 corresponding to the guiding center motions,
 are generated in Eq.~\ref{eqno1},
 as will be addressed in Sec.3.

With the surface charges having ready access to the field lines,
 plasma density will be high enough in the magnetosphere
 such that plasma conductivity $\sigma$ tends to be infinite,
 the Ohm law reads

\begin{eqnarray}
\label{eqno2}
\vec E + \vec v\times\vec B\,
 =\,{1\over\sigma}\vec J\,
 =\,0\,\,\,.
\end{eqnarray}

\noindent For a quasi-neutral normal
 magnetohydrodynamic (MHD) plasma,
 the plasma velocity $\vec v$ in Eq.~\ref{eqno2}
 amounts to the center of mass velocity.
 For a charge separated plasma,
 this equation coincides with Eq.~\ref{eqno1}
 other than an overall fator
 of the space charge density $\rho_{q}$.
 Nevertheless, this does not mean that Eq.~\ref{eqno2} is redundant,
 since it describes the magnetosphere in terms of velocities
 while Eq.~\ref{eqno1} does it in terms of fields.
 These two descriptions are then coupled
 through the space charge densities.
 
Probably induced by the lighthouse scenario
 of pulsar beacons,
 the pulsar magnetosphere is often interpreted
 as such that the magnetic field lines
 are rigidly anchored on the pulsar surface.
 The entire magnetosphere thus corotates
 with the pulsar at its angular velocity,
 and the charge separated plasma
 is being dragged along with it.
 Although for an aligned rotator and under axisymmetry,
 it is indistinguishable whether or not
 the magnetosphere is rotating.
 Nevertheless, in an oblique rotator,
 it could generate the lighthouse beacons
 and magnetic dipole radiation.
 In this scenario,
 the rotating magnetic field $\vec B$ drags the plasma
 generating a rigid rotor velocity $\vec v$.
 Applying this rotating plasma velocity $\vec v$
 back to a stationary frame in space,
 and considering the magnetic field $\vec B$ be stationary,
 which is indistinguishable from a rotating one,
 Eq.~\ref{eqno2} then gives
 the magnetospheric electric field $\vec E$
 in the stationary frame.
 We remark that only under the axisymmetric aligned rotator case
 that the magnetic field can be considered
 stationary and/or rotating by option.
 So the plasma velocity $\vec v$
 is an input parameter in Eq.~\ref{eqno2},
 the driving energy is the rotating magnetic field,
 and the electric field $\vec E$ is the response
 which is called the rotation induced electric field.
 Some authors introduce a non-rotation induced
 longitudinal field $-\nabla V$
 by taking into consideration
 of the interstellar plasma potential,
 or by considering the centrifugal outflow
 at the light cylinder (LC)
 \citep{mestel1985, fitzpatrick1988, goodwin2004, timokhin2006}.
 Other authors have considered a very stringent supply
 of positive charges from the pulsar surface,
 due to probably the surface physics of neutron star,
 and proposed that vacuum gaps could be formed
 separating the charged regions
 from the pulsar surface
 and/or separating the charged regions
 from each other
 \citep{ruderman1975, jackson1976, michel1979}.

In this corotating magnetic field model,
 the unipolar induction in the neutron star
 is clearly decoupled from the magnetosphere.
 The magnetic dipole radiation,
 the angular momentum outflow of the stellar winds,
 and the current loop in the presence of $-\nabla V$
 are attributed to the braking torque on the pulsar indirectly
 through the general energy conservation principle.
 However, the use of the energy conservation principle
 implies a direct coupling with the neutron star,
 which is not reflected in this scenario.
 As a matter of fact,
 if the magnetosphere field lines
 are anchored on the pulsar surface,
 the (dipole) magnetic field inside the pulsar
 will be rotating as well.
 Consequently, there is no relative rotation
 between the pulsar and the magnetic field inside.
 For this reason, there will be no unipolar induction as well.

As a matter of fact,
 a steady state magnetosphere decoupled from the pulsar
 implies that the emission problem
 of the beacons \citep{melrose1978}
 can be constructed from the equilibrium magnetosphere.
 Although the validity
 of separating equilibrium and emission
 into two different issues
 cannot be disproved,
 this approach could have over simplified
 the pulsar dynamics. 
 Recently, a time dependent description of magnetosphere
 where the plasma equations are coupled
 to the Maxwell equations
 to account for induction effects.
 This broader approach has been adopted
 to study the formation of magnetosphere current sheets
 \citep{komissarov2006, spitkovsky2006, tchekhovskoy2013},
 and the linear instabilities in magnetosphere
 that could be responsable for emissions
 \citep{urpin2012, urpin2014}.

\newpage
\section{Neutron Star-Stationary Magnetosphere Coupled System}

To overcome the rotation ambiguities
 of plasma velocity and magnetic field, 
 we consider the magnetic field $\vec B$ be stationary in space,
 and the electric field $\vec E$
 is uploaded to the equipotential field lines
 of the magnetosphere
 according to the potential distribution
 on the pulsar surface
 estblished by the unipolar induction.
 This means that the charge separated magnetospheric plasma
 has a negligible contribution to the electric field there.
 In this case of stationary magnetic field $\vec B$,
 the driving energy is the uploaded electric field $\vec E$
 which is the input parameter in Eq.~\ref{eqno2},
 and the plasma velocity $\vec v$ is the response.
 The magnetosphere is directly coupled
 to the unipolar induction of the pulsar.
 The charge separation flow inside the pulsar
 generates a current.
 The interaction of this current with the magnetic field
 produces a direct braking torque
 in addition to the contributions
 of dipole radiation and stellar winds
 which slow down the rotation rate.
 Finally, we remark that magnetic field
 is a real physical quantity,
 while magnetic field line
 is only a physical concept
 derived from the field line equation
 to help visualize the action of magnetic field.
 It is rather unphysical to consider field lines
 being rigidly attached to the pulsar surface
 corotating with it.
 The notion of a rotating magnetosphere
 is probably cultivated in the oblique rotator case,
 where the magnetosphere is seen to rotate
 obliquely about the pulsar rotational axis.
 As obliqueness goes to zero,
 we then have an aligned rotator.
 For an oblique rotator,
 we choose to phrase the magnetosphere
 as wobbling (not rotating) about the rotational axis.
 Consequently, as the obliqueness goes to zero,
 the magnetosphere stops to wobble
 and becomes stationary in space.
 
Here, in this paper, we reconsider 
 the \citet{goldreich1969} analysis
 of charge separated plasmas.
 In particular, there are two inconsistencies in their work.
 The first inconsistency is the space charge density expression
 which is incompatible with the curvature
 of the closed ion magnetic field lines near the pulsar.
 To resolve this inconsistency,
 they resorted to a corotating electron cloud
 above the closed magnetic field lines.
 The second one is the introduction
 of the far zone quasi-neutral interstellar plasma potential
 to the near zone pulsar magnetosphere
 to draw plasma outflows in the open field lines.
 The field line that has the potential
 of the interstellar plasma
 is the critical field line
 with no charge flow.
 This interstellar emf is inconsistent
 to the $E_{\parallel}=0$ plasma condition
 in the magnetosphere.
 Actually, many publications in recent decades
 have concentrated on the pulsar equation itself
 without mentioning
 neither the question of space charge density
 nor the condition of the critical field line.
 These simplifications are equivalent
 to considering the pulsar magnetosphere
 as filled with a quasi-neutral normal plasma
 instead of a charge separated plasma.

The charge separated plasma
 in this stationary magnetosphere
 undergoes an $\vec E\times\vec B$ drift
 which happens to have an angular velocity of the pulsar,
 instead of being dragged along
 by the corotating magnetosphere.
 As for the plasma outflow,
 we consider it be driven by plasma drifts,
 not by interstellar emf.
 Finally, with a stationary magnetosphere in space,
 coupled to the unipolar induction in the pulsar interior,
 we offer an alternative pulsar beacon mechanism
 by suggesting a pulsating magnetosphere.
 To understand this dynamic picture,
 we assume that the unipolar induction
 is delivering more power
 to the magnetosphere
 than the open field lines can drain.
 This leads to an energy built-up
 in the closed field lines
 represented by plasma pressure.
 Although plasma pressure is neglected
 in Eq.~\ref{eqno1},
 it appears indirectly through plasma drifts,
 which could open up the closed field lines near LC.

Observationally, radio pulsars are known
 for their stable repetitive pulse profiles.
 However, the pulse profile
 that characterizes each radio pulsar
 is obtained as the average over many many shots.
 From shot to shot, the profiles are far from identical.
 Therefore, the period of oblique rotator
 is not well defined from shot to shot,
 which is not supposed to be in the lighthouse scenario.
 On the other hand, such shot to shot variations
 are expected in a pulsating magnetosphere.

We review the basic concepts
 of charge separated pulsar magnetosphere
 and the plasma velocity $\vec v$ of Eq.~\ref{eqno2} in Sec.3.
 The space charge density $\rho_{q}$
 is derived in the presence of a poloidal and toroidal field,
 and in the presence of drift velocities in Sec.4.
 Under the concept of guiding center drifts,
 the idea of a critical field line
 is reconsidered in Sec.5.
 We then follow the earlier works of
 \citet{sturrock1971},
 \citet{scharlemann1973},
 \citet{okamoto1974},
 \citet{beskin1998},
 \citet{goodwin2004},
 to solve the magnetosphere
 of an isolated pulsar analytically
 via the force-free pulsar equation.
 Nevertheless, we take a different approach
 to split the pulsar equation
 into two coupled equations
 through a separating function.
 Each one of the coupled equations
 is solved by separation of variables
 for the global structure
 of the pulsar magnetosphere in Sec.6.
 The standard monopole-like
 steady state magnetosphere
 is presented in Sec.7.
 A self-consistent description of magnetosphere
 in terms of fields, velocities,
 and space charge densities
 is outlined in Sec.8.
 Under continuous unipolar pumping,
 the standard magnetosphere
 is generalized to the pulsating magnetosphere in Sec.9,
 and with conclusions in Sec.10.

\newpage
\section{Plasma Velocity}

It is important to remark
 that Eq.~\ref{eqno2} is written in an inertial frame
 where the plasma moves with velocity $\vec v$
 with respect to the stationary
 magnetic and electric fields.
 In the plasma moving frame,
 the Lorentz transformed electric field
 is $\vec E'=\vec E+\vec v\times\vec B$
 and the Ohm law reads $\vec E'=0$ there.
 This Ohm law in the limit of infinite $\sigma$
 is known as the frozen field condition
 in MHD plasmas,
 and happens to be identical to Eq.~\ref{eqno1},
 and we have $\vec E\cdot\vec v=0$
 and $E_{\parallel}=\vec E\cdot\vec B=0$.
 As a matter of fact,
 for a charge separated plasma,
 Eq.~\ref{eqno1} and Eq.~\ref{eqno2}
 amount to the same equation always.
 However, Eq.~\ref{eqno1}
 describes the magnetosphere in terms of fields,
 while Eq.~\ref{eqno2}
 considers the magnetosphere in terms of plasma velocities.
 The fields and the velocities
 are then connected through the space charge density.

Here we call to the attention
 that the condition of $E_{\parallel}=0$
 allows an arbitrary longitudinal velocity
 $\vec v_{\parallel}$ along the magnetic field line.
 Consequently, the plasma velocity
 can be projected in components
 parallel and perpendicular
 to the magnetic field as

\begin{eqnarray}
\nonumber
\vec v\,
 =\,\vec v_{\parallel}+\vec v_{\perp}\,\,\,,
\\
\nonumber
\vec v_{\perp}\,=\,\vec v_{E}\,
 =\,{1\over B^{2}}(\vec E_{\perp}\times\vec B)\,\,\,.
\end{eqnarray}

\noindent The frozen field plasma condition
 shows that $\vec v_{E}$ is given by the $\vec E\times\vec B$
 guiding center drift of the gyro motion
 due to the transverse part of the electric field $\vec E$
 \citep{thompson1962, schmidt1966, jackson1975},
 under the constrain of $v_{E}<c$
 which implies the electric field strength be
 $E_{\perp}<cB$ for a magnetized plasma.
 This $\vec v_{E}$ drift is charge independent,
 and for a quasi-neutral normal plasma,
 this drift generates no current
 and represents only the fluid drift velocity.
 But for charge separated plasmas,
 the $\vec E\times\vec B$ drift corresponds a current flow.
 Consequently, we have the picture
 of a stationary pulsar magnetosphere
 with plasma drifting
 with respect to the magnetic field.

In cylindrical coordinates $(r,\phi,z)$,
 the axisymmetric magnetic field can be represented as

\begin{eqnarray}
\label{eqno3}
\vec B\,=\,A_{0}(\nabla\Psi\times\nabla\phi+\alpha A\nabla\phi)\,
 =\,A_{0}\left(-{1\over r}{\partial\Psi\over \partial z},
 +{1\over r}\alpha A, +{1\over r}{\partial\Psi\over\partial r}\right)
 \,\,\,,
\\
\nonumber
\mu_{0}\vec J\,=\,A_{0}{1\over r}
 \left(-\alpha{\partial A\over\partial z},
 -\nabla^{2}\Psi+{2\over r}{\partial\Psi\over\partial r},
 +\alpha{\partial A\over\partial r}\right)
 \,\,\,,
\end{eqnarray}

\noindent where $A_{0}$ carries the dimension
 of the poloidal magnetic flux
 such that $\Psi$ is a dimensionless flux function,
 and $\alpha=A_{\phi}/A_{0}$ represents the relative ratio
 of the toroidal flux to the poloidal flux.
 The function $A$ is related to the axial current through

\begin{eqnarray}
\label{eqno4}
2\pi\alpha A_{0}A\,=\,\mu_{0}I_{z}\,\,\,.
\end{eqnarray}

\noindent From the magnetic field line equation,
 it is easy to show that the poloidal field lines
 are given by the contours of $\Psi$

\begin{eqnarray}
\label{eqno5}
\Psi(r,z)\,=\,C\,\,\,.
\end{eqnarray}

\noindent The fact that the longitudinal electric field vanishes
 means that any potential difference
 along the magnetic field line
 will be shorted out by plasma mobility.
 With the longitudinal electric field $E_{\parallel}=0$
 and the magnetic field $\vec B$
 represented by Eq.~\ref{eqno3},
 the transverse electric field in the magnetosphere
 can be expressed through the equipotential field lines
 of Eq.~\ref{eqno5} as

\begin{eqnarray}
\nonumber
\vec E_{\perp}\,=\,-V_{0}\nabla\Psi\,\,\,,
\end{eqnarray}

\noindent where $V_{0}$ is the equator-pole voltage drop
 on the pulsar surface,
 and $\Psi$ is assigned by the corresponding
 surface potential label.

Alternatively, we can also project the plasma velocity
 in poloidal and toroidal components
 by writing
 $\vec v=\vec v_{p}+\vec v_{\phi}$ and
 $\vec B=\vec B_{p}+\vec B_{\phi}$.
 From Eq.~\ref{eqno2}, we get
 $0=-\vec E_{\phi}=\vec v_{p}\times\vec B_{p}$ by axisymmetry,
 and
 $-\vec E_{p}=\vec v_{p}\times\vec B_{\phi}
 +\vec v_{\phi}\times\vec B_{p}$.
 We therefore have respectively

\begin{eqnarray}
\label{eqno6}
\vec v_{p}\,=\,\kappa_{v}\vec B_{p}\,\,\,,
\\
\label{eqno7}
-\vec E_{p}\,
 =\,(-\kappa_{v}\vec B_{\phi}+\vec v_{\phi})
 \times\vec B_{p}\,\,\,.
\end{eqnarray}

\noindent Loading the potential distribution
 of the pulsar surface onto the equipotential
 magnetosphere poloidal field lines,
 we have 

\begin{eqnarray}
\nonumber
-\vec E_{p}\,
 =\,r\Omega_{0}\hat\phi\times\vec B_{p}\,\,\,,
\end{eqnarray}

\noindent where $r$ measures the radial position
 of a point $\vec r$ in the magnetosphere.
 Substituting this result to Eq.~\ref{eqno7}
 then gives
 $\vec v_{\phi}=\kappa_{v}\vec B_{\phi}+r\Omega_{0}\hat\phi$,
 and therefore

\begin{eqnarray}
\label{eqno8}
\vec v\,=\,\vec v_{p}+\vec v_{\phi}\,
 =\,\kappa_{v}\vec B+r\Omega_{0}\hat\phi\,\,\,.
\end{eqnarray}

\noindent We call to the attention
 that $\vec v_{\phi}$ drifts
 with an angular velocity of the pulsar.
 This equation is often interpreted
 as the 'beads on a wire' model,
 where the plasma (beads) slides
 with a velocity $\kappa_{v}\vec B_{\phi}$
 along a rotating $r\Omega_{0}\hat\phi$ field line (wire).
 We have rederived this equation
 with a stationary magnetosphere
 where the rotation comes from the guiding center drift.

We have expressed the plasma velocity
 in two orthogonal projections.
 The first one is with respect to the magnetic field lines,
 and the second one is with respect to
 the cylindrical coordinate system.
 These two projections are independent to each other.
 Only when the magnetic field is purely poloidal
 with $B_{\phi}=0$
 that the two expressions are correspondent,
 and we can identify

\begin{eqnarray}
\nonumber
\vec v_{\parallel}\,=\,\vec v_{p}\,
 =\,\kappa_{v}\vec B_{p}\,\,\,,
\\
\nonumber
\vec v_{\perp}\,=\,\vec v_{\phi}\,=\,\vec v_{E}\,
 =\,r\Omega_{0}\hat\phi\,\,\,.
\end{eqnarray}

\newpage
\section{Space Charge Density}

In the magnetosphere,
 the electric field $\vec E$
 is not the only transverse force
 that causes plasma drift.
 Magnetic field line curvature and field gradient
 are other two major transverse forces.
 Therefore, besides the $\vec v_{E}$ drift,
 we also have the curvature and gradient B drifts
 \citep{thompson1962, schmidt1966, jackson1975}
 given by

\begin{mathletters}
\begin{eqnarray}
\label{eqno9a}
\vec v_{Rc}\,
 =\,+{2\epsilon_{\parallel}\over q}
 {1\over B^{2}}
 ({\vec R_{c}\over R_{c}^{2}}\times\vec B)
 \,\,\,,
\\
\label{eqno9b}
\vec v_{\nabla B}\,
 =\,-{\epsilon_{\perp}\over q}
 {1\over B^{2}}
 ({\nabla B\over B}\times\vec B)
 \,\,\,,
\end{eqnarray}
\end{mathletters}

\noindent where $\epsilon_{\parallel}=mv_{\parallel}^{2}/2$
 and $\epsilon_{\perp}=mv_{\perp}^{2}/2$
 are the charge particle energies
 parallel and perpendicular to the magnetic field.
 Therefore, these drifts are mass dependent
 as well as field dependent.
 In a thermal plasma,
 these energies are measued by the plasma pressure.
 Although plasma inertia and pressure
 are neglected in Eq.~\ref{eqno1},
 they appear indirectly through these two drift velocities.
 The direction of these two drifts
 depends on the sign of charge q.
 Even in a quasi-neutral normal plasma,
 these are current generating drifts,
 while the electric drift $\vec v_{E}$
 represents the plasma fluid velocity only.
 In a charge separated plasma,
 all three drifts generate currents.
 These plasma drifts are driven by transverse forces,
 and they are not emf driven velocities.
 To understand this,
 we recall that for a uniform magnetic fleld,
 charge particles follow a circular gyro motion
 along the magnetic field lines
 according to the equation of motion
 due to the plasma inertia.
 In the presence of a transverse force to the magnetic field
 such as an electric field, a centrifugal force due to curvature,
 and an inhomogeneous magnetic field, etc.,
 the circular gyro motion will be transformed
 into a displacing cycloid across the field lines
 \citep{thompson1962, schmidt1966, jackson1975},
 such that the gyro guiding center
 presents a transverse drift.
 Although Eq.~\ref{eqno2} gives the guiding center
 electric drift directly,
 the gyro motion itself is due to the plasma inertia.
 Therefore, Eq.~\ref{eqno2} actually
 includes the plasma inertia implicitly.
 We remark that these drift velocities
 can be very large and become relativistic.
 Should the electric field $\vec E$ be time varying,
 there would be a polarization drift as well
 which we will ignore here.

We should point out that in Eq.~\ref{eqno2}
 only the $\vec v_{E}$ drift is represented on the left side.
 The other two current generating drifts,
 that are contained in $\vec J$
 on the right side of Eq.~\ref{eqno2},
 do not appear in Eq.~\ref{eqno2}
 because the plasma conductivity $\sigma$ is infinite.
 For this reason, the plasma drift velocity is given by

\begin{eqnarray}
\nonumber
\vec v_{drift}\,
 =\,(\vec v_{Rc}+\vec v_{\nabla B})+\vec v_{E}\,\,\,.
\end{eqnarray}

\noindent Since the plasma drift velocities
 are obtained from the equation of motion
 of each charged plasma,
 the plasma drift velocity $\vec v_{drift}$
 is beyond the scope of Eq.~\ref{eqno2},
 as discussed in Sec.1,
 in the sense that the MHD expression of
 $\vec v=\vec v_{\parallel}+\vec v_{E}$
 differs from $\vec v_{drift}$
 because 
 $\vec v_{\parallel}\neq (\vec v_{Rc}+\vec v_{\nabla B})$.

The space charge density,
 which connects the fields of Eq.~\ref{eqno1}
 and the velocities of Eq.~\ref{eqno2},
 in the magnetosphere is, by Eq.~\ref{eqno2},

\begin{eqnarray}
\nonumber
\rho_{q}\,
 =\,-\epsilon_{0}\nabla\cdot(\vec v_{E}\times\vec B)\,
 =\,\epsilon_{0}(\vec v_{E}\cdot\mu_{0}\vec J
 -\vec B\cdot\nabla\times\vec v_{E})\,\,\,.
\end{eqnarray}

\noindent Writing
 $\vec J=\rho_{q}\vec v_{drift}
 =\rho_{q}(\vec v_{E}+\vec v_{Rc}+\vec v_{\nabla B})$,
 considering a general poloidal and toroidal fields,
 and representing $\vec v_{E}$ by the projection of Eq.~\ref{eqno8}
 to evaluate the $-\vec B\cdot\nabla\times\vec v_{E}$ term
 on the right side, we get
 
\begin{eqnarray}
\label{eqno10}
[\rho_{q}-\epsilon_{0}\mu_{0}(\vec v_{E}-\kappa_{v}\vec B)\cdot\vec J]\,
 =\,[1-\epsilon_{0}\mu_{0}(\vec v_{E}-\kappa_{v}\vec B)
 \cdot\vec v_{drift}]\rho_{q}\,
 =\,-2\epsilon_{0}\vec B\cdot\vec\Omega_{0}\,\,\,.
\end{eqnarray}

\noindent In the particular case of a purely poloidal field,
 we can identify $\vec v_{E}=r\Omega_{0}\hat\phi$,
 and Eq.~\ref{eqno10} becomes
 
\begin{eqnarray}
\nonumber
(\rho_{q}-\epsilon_{0}\mu_{0}\vec v_{E}\cdot\vec J)\,
 =\,(1-\epsilon_{0}\mu_{0}\vec v_{E}\cdot\vec v_{drift})\rho_{q}\,
 =\,-2\epsilon_{0}\vec B\cdot\vec\Omega_{0}\,\,\,.
\end{eqnarray}

\noindent \citet{goldreich1969}
 had derived a space charge density (their Eq.8)
 with $\vec J=\rho_{q}\vec v_{E}$ alone that reads

\begin{eqnarray}
\nonumber
(1-\epsilon_{0}\mu_{0}v_{E}^{2})\rho_{q}\,
 =\,-2\epsilon_{0}\vec B\cdot\vec\Omega_{0}\,\,\,.
\end{eqnarray}

\noindent In the absence of $\vec v_{Rc}$ and $\vec v_{\nabla B}$,
 the left side of the above equation
 looks like the relativistic Lorentz factor,
 which is just coincidental.
 This equation predicts $B_{z}<0$ for closed ion field lines,
 which is consistent only to the outer portion
 of the dipole-like closed ion field lines.
 For the inner portion with $B_{z}>0$,
 they postulated a negative charged cloud
 corotating with the field lines.

We antecipate the result in Sec.8 that $\kappa_{v}<0$,
 such that the $(\vec v_{E}-\kappa_{v}\vec B)$ term is positive.
 Near the pulsar, $\vec v_{Rc}$ and $\vec v_{\nabla B}$ are large,
 where the plasma pressure is very high,
 on the left side
 and we could have $B_{z}>0$.
 Farther out from the pulsar
 with plasma pressure decreasing
 we have $B_{z}<0$.
 Since $\vec v_{Rc}$ and $\vec v_{\nabla B}$ drifts
 are mass dependent,
 they would be dominant again
 as the closed field lines approach LC,
 due to their relativistic inertia.
 We would have $B_{z}>0$ on approaching LC.
 For this reason,
 the closed region of a steady state magnetosphere
 can only extend to some intermediate radius,
 staying away from LC.
 Although plasma pressure and plasma inertia
 are not considered in Eq.~\ref{eqno1} explicitly,
 plasma inertia appears implicitly
 through the guiding center drifts of the gyro motion
 and plasma pressure through the coefficients
 of the gradient B and curvature drifts.
 We have given a qualitative understanding
 of these effects in the closed region
 through the drift velocities,
 and we will return to this issue in Sec.8.

\newpage
\section{Critical Field Line}

In terms of cylindrical coordinates,
 the toroidal components of Eq.~\ref{eqno1},
 which describes the magnetosphere in terms of fields,
 gives

\begin{eqnarray}
\nonumber
(\alpha\nabla A\times\nabla\phi)\times(\nabla\Psi\times\nabla\phi)\,
 =\,0\,\,\,,
\\
\label{eqno11}
A\,=\,A(\Psi)\,\,\,,
\end{eqnarray}

\noindent whereas the poloidal components
 yield the pulsar equation

\begin{eqnarray}
\label{eqno12}
\left(1-({r\over r_{L}})^{2}\right)\nabla^{2}\Psi
 -{2\over r}{\partial\Psi\over\partial r}
 +\alpha^{2} A(\Psi){\partial A(\Psi)\over\partial\Psi}\,
 =\,0\,\,\,,
\end{eqnarray}

\noindent where $r_{L}=c/\Omega_{0}$ is the LC radius.
 Because of the singular coefficient
 of the highest derivative
 which vanishes at $r/r_{L}=1$,
 Eq.~\ref{eqno12} is a singular equation.
 As indicated by \citet{scharlemann1973} in Sec.III,
 the steady state magnetosphere,
 which is stationary in space, is solved within LC,
 and is then analytically continued beyond LC
 where the pulsar equation is not valid
 with the tengential velocity $v_{\phi}=r\Omega_{0}>c$.
 Since plasma parameters
 like plasma pressure and velocities
 are not considered in the pulsar equation,
 analytic continuation of the field structure beyond LC
 amounts to solving the pulsar equation in that region
 with solutions on both sides matched across LC,
 as done numerically by
 \citet{contopoulos1999},
 \citet{timokhin2006}, and others.
 
Analytic solutions of this equation
 had been extensively studied,
 yet global solutions were not obtained explicitly
 \citep{scharlemann1973, okamoto1974}.
 So far, solutions are developed by numerical iterrations,
 under the boundary condition of a dipole field at the center
 notably by
 \citet{contopoulos1999},
 \citet{ogura2003},
 \citet{gruzinov2005},
 \citet{contopoulos2005}
 where the open and closed regions rotate differentially,
 and \citet{timokhin2006}
 where the braking index is evaluated
 in terms of the magnetosphere dynamics.
 
In the open field lines, electrons and ions are assummed
 to stream out along their respective
 open field lines with equal flux.
 These currents will be connected
 to form a return loop at a distant load (interstellar plasma).
 The electron and ion outflows
 are thought to be drawn
 by the action of the far zone quasi-neutral
 normal interstellar plasma floating potential
 $\vec E_{inter}=\pm V_{inter}\nabla\Phi$,
 where $V_{inter}$ is the potential difference
 between the far zone interstellar plasma
 and the near zone electron or ion open field lines,
 and $\Phi$ is a dimensionless scalar potential function.
 In particular, there will be a field line
 having the same potential of the interstellar plasma.
 This is the critical field line $\Psi_{c}$,
 where there is no charge and no poloidal current flow,
 that separates the electron and ion field lines.
 This means that $\partial A(\Psi)/\partial\Psi=0$,
 and $A(\Psi)$ should be stationary at $\Psi_{c}$.
 This critical field line
 should pierce LC with $B_{z}=0$.

Nevertheless, using the interstellar emf
 to draw current flows in the open field lines
 is clearly in confrontation
 with the $E_{\parallel}=0$ condition.
 This inconsistency can be removed
 should we note that the presence of a toroidal
 and a poloidal magnetic field in the open region
 can drive a poloidal and a toroidal
 $\vec v_{E}$ drift respectively
 
\begin{eqnarray}
\nonumber
\vec v_{E}\,
 =\,\vec v_{p}+\vec v_{\phi}\,
 =\,{1\over B^{2}}
 \vec E_{\perp}\times(\vec B_{\phi}+\vec B_{p})\,\,\,.
\end{eqnarray}

\noindent The same can be applied to
 $\vec v_{Rc}$ and $\vec v_{\nabla B}$ drifts.
 All these drift velocities will, therefore,
 produce a component along the magnetic field lines
 plus a toroidal rotation, as in Eq.~\ref{eqno8},
 and there is no need to introduce
 neither the interstellar potential to the near zone
 nor the concept of a critical field line.
 
\newpage
\section{Analytic Pulsar Solution}

By considering the magnetic field $\vec B$
 be stationary in space,
 and the electric field $\vec E$
 be uploaded to the field lines from the pulsar surface,
 our model has the plasma velocity $\vec v$
 as the response in Eq.~\ref{eqno2}.
 With $\vec v_{\phi}=r\Omega_{0}\hat\phi$,
 our model simply implies
 that the electric potential on the pulsar surface
 cannot be fully uploaded to the magnetosphere
 to all distances, even beyond LC.
 This means that the plasma conductivity $\sigma$ is finite
 in Eq.~\ref{eqno2} on approaching LC,
 and collisional effect has to be taken into account,
 or equivalently, inertia effect.
 Likewise is in Eq.~\ref{eqno1}.
 This should be the subject of further investigations.
 For the moment, we persuit Eq.~\ref{eqno12} as it is.

To construct the standard magnetosphere,
 we denote $\xi=r/r_{L}$ as the normalized radial coordinate
 and take a different approach
 to recast the pulsar equation as

\begin{eqnarray}
\label{eqno13}
\nabla^{2}\Psi\,
 =\,{1\over (1-\xi^{2})}
 \left[{1\over r_{L}^{2}}{2\over\xi}{\partial\Psi\over\partial\xi}
 -\alpha^{2} A(\Psi){\partial A(\Psi)\over\partial\Psi}\right]\,
 =\,f(\Psi)\,\,\,,
\end{eqnarray}

\noindent where we have required both sides of the equality
 be equal to a separating function $f(\Psi)$.
 We attempt to solve this equation analytically
 by choosing

\begin{eqnarray}
\label{eqno14}
f(\Psi)\,=\,k^{2}\Psi\,\,\,,
\end{eqnarray}

\noindent which renders

\begin{mathletters}
\begin{eqnarray}
\label{eqno15a}
{\partial^{2}\Psi\over\partial\xi^{2}}
 +{1\over\xi}{\partial\Psi\over\partial\xi}
 +{\partial^{2}\Psi\over\partial\varsigma^{2}}\,
 =\,(kr_{L})^{2}\Psi\,\,\,,
\\
\label{eqno15b}
{2\over\xi}{\partial\Psi\over\partial\xi}
 -(\alpha r_{L})^{2}A(\Psi){\partial A(\Psi)\over\partial\Psi}\,
 =\,(1-\xi^{2})(kr_{L})^{2}\Psi\,\,\,,
\end{eqnarray}
\end{mathletters}

\noindent where $\varsigma=z/r_{L}$ is the normalized axial coordinate.
 We write $\Psi(\xi,\varsigma)=R(\xi)Z(\varsigma)$
 by separation of variables,
 and Eq.~\ref{eqno15a} gives

\begin{mathletters}
\begin{eqnarray}
\label{eqno16a}
{\partial^{2}Z\over\partial\varsigma^{2}}
 -((kr_{L})^{2}-m^{2})Z\,
 =\,0\,\,\,,
\\
\label{eqno16b}
{\partial^{2}R\over\partial\xi^{2}}
 +{1\over\xi}{\partial R\over\partial\xi}\,
 =\,m^{2}R\,\,\,,
\end{eqnarray}
\end{mathletters}

\noindent where $m^{2}$ is the separation constant.
 We have denoted the two constants
 in Eq.~\ref{eqno14} and Eq.~\ref{eqno16b}
 by $k^{2}$ and $m^{2}$ respectively
 in quadratic form by choice.
 However, $k^{2}$ and $m^{2}$
 are not necessarily be positive.
 As for Eq.~\ref{eqno15b}, we take a linear $A(\Psi)$ and get

\begin{mathletters}
\begin{eqnarray}
\label{eqno17a}
A(\Psi)\,=\,-k_{z}\Psi\,\,\,,
\\
\label{eqno17b}
{2\over\xi}{\partial R\over\partial\xi}\,
 =\,[(\alpha k_{z}r_{L})^{2}+(kr_{L})^{2}(1-\xi^{2})]R\,\,\,.
\end{eqnarray}
\end{mathletters}

\noindent Here, $k_{z}$ has the dimension
 of an inverse scale length.
 We have chosen a minus sign explicitly
 in Eq.~\ref{eqno17a} to represent an overall
 downward axial current in the northern polar region.
 With the understanding that electrons and ions
 flow along their field lines by plasma drifts,
 $A(\Psi)$ does not have to have a stationary point on LC
 which ables the linear choice in Eq.~\ref{eqno17a}.
 Such choice of $A(\Psi)=-k_{z}\Psi$
 was also used by \citet{scharlemann1973}
 and \citet{beskin1998}.
 Our analytic approach to the pulsar equation
 has introduced a separating function $f(\Psi)=k^{2}\Psi$,
 a separating constant $m^{2}$,
 and an axial current function $A(\Psi)=-k_{z}\Psi$.

The $Z(\varsigma)$ function can be solved readily to render

\begin{mathletters}
\begin{eqnarray}
\label{eqno18a}
Z_{a}(\varsigma)\,
 =\,C_{a}e^{-\kappa\varsigma}\,\,\,,
\\
\label{eqno18b}
Z_{b}(\varsigma)\,
 =\,C_{b}e^{+\kappa\varsigma}\,\,\,,
\end{eqnarray}
\end{mathletters}

\noindent with $\kappa^{2}=((kr_{L})^{2}-m^{2})=(a_{L}^{2}-m^{2})>0$,
 and

\begin{mathletters}
\begin{eqnarray}
\label{eqno19a}
Z'_{a}(\varsigma)\,
 =\,C'_{a}\cos{\kappa'\varsigma}\,\,\,,
\\
\label{eqno19b}
Z'_{b}(\varsigma)\,
 =\,C'_{b}\sin{\kappa'\varsigma}\,\,\,,
\end{eqnarray}
\end{mathletters}

\noindent with $\kappa^{2}=-\kappa'^{2}=-(m^{2}-a_{L}^{2})<0$.
 To solve for $R(\xi)$, we make use of Eq.~\ref{eqno17b}
 to cast Eq.~\ref{eqno16b} as

\begin{eqnarray}
\nonumber
2{\partial^{2}R\over\partial\xi^{2}}
 +[((\alpha k_{z}r_{L})^{2}+(kr_{L})^{2}-2m^{2})
 -(kr_{L})^{2}\xi^{2}]R\,
\\
\nonumber
 =\,2{\partial^{2}R\over\partial\xi^{2}}
 +[(a_{z}^{2}+a_{L}^{2}-2m^{2})
 -a_{L}^{2}\xi^{2}]R\,
\\
\label{eqno20}
 =\,2{\partial^{2}R\over\partial\xi^{2}}
 +[a_{\phi}^{2}
 -a_{L}^{2}\xi^{2}]R\,
 =\,0\,\,\,.
\end{eqnarray}

\noindent This equation can be solved through a power series.
 Because of the boundary condition of $B_{\phi}(r,z)=0$
 on the polar axis $r=0$,
 we should have $R(\xi)=\xi^{2}$ as $\xi$ goes to zero.
 We therefore have

\begin{eqnarray}
\label{eqno21}
R(\xi)\,=\,\xi^{2}\Sigma a_{n}\xi^{n}\,\,\,,
\\
\nonumber
a_{2}\,=\,-{1\over 24}a_{\phi}^{2}a_{0}\,\,\,,
\\
\nonumber
a_{3}\,=\,-{1\over 40}a_{\phi}^{2}a_{1}\,\,\,,
\\
\label{eqno22}
a_{n}\,=\,-{1\over 2(n+2)(n+1)}
 (a_{\phi}^{2}a_{n-2}-a_{L}^{2}a_{n-4})\,\,\,.
\end{eqnarray}

\noindent We note that $a_{0}$ and $a_{1}$
 are the two independent constants
 of the second order differential equation, Eq.~\ref{eqno20}.
 Other coefficients are generated
 by the recurssion formula, Eq.~\ref{eqno22}.
 The coefficients $a_{0}$ and $a_{1}$
 generate even and odd power terms
 of the series respectively.
 Thus $a_{0}$ and $a_{1}$ are the multipliers
 of the even and odd series
 that we can write

\begin{eqnarray}
\label{eqno23}
R(\xi,a_{0},a_{1})\,
 =\,a_{0}R_{even}(\xi,1,0)+a_{1}R_{odd}(\xi,0,1)\,\,\,.
\end{eqnarray}

\newpage
\section{Standard Magnetosphere}

We note that the radial function $R(\xi)$
 is governed by the parameters $(a_{\phi}^{2},\,a_{L}^{2})$
 and the choice of the two independent coefficients
 $(a_{0},\,a_{1})$.
 With positive $a_{\phi}^{2}$ and $a_{L}^{2}$,
 the coefficients $a_{n}$ are alternating in sign,
 and $R(\xi)$ could have oscillating solutions for small $\xi$.
 This functional form is suitable to construct
 the open ion field lines and the closed region
 of the pulsar magnetosphere.
 As $\xi$ increases,
 $R(\xi)$ becomes monotonically increasing,
 which is suitble for the split monopole structure
 far from the pulsar.
 As for the axial function $Z_{a}(\varsigma)$,
 it is a decreasing function of $\varsigma$.
 When this is combined with the radial function,
 together they could generate contours
 adequate for open field lines.
 For a given $a_{\phi}^{2}$,
 we could scan $a_{L}^{2}$ over a given range
 to generate a set of field line morphologies.
 As a result, we represent
 the electron open field lines
 by taking the poloidal flux function as

\begin{eqnarray}
\label{eqno24}
\Psi(\xi,\varsigma)\,
 =\,R(\xi)Z_{a}(\varsigma)\,
 =\,C\,\,\,.
\end{eqnarray}

\noindent For simplicity,
 we consider the even terms of the power series
 by taking $a_{0}=1$ and $a_{1}=0$.
 We choose to set $a_{\phi}^{2}=20$, say,
 for the radial solution.
 Taking $a_{L}^{2}=8$,
 the function $R(\xi)$ is a monotonically increasing function,
 as is shown in Fig.1
 which is appropriate for electron open field lines.
 We should emphasize that the monotonic
 functional form of $R(\xi)$
 can be obtained under a wide range of parameters,
 not restricted to the chosen ones.
 Furthermore, we set $\kappa^{2}=5$, for example,
 and we have the self-consistent parameters
 $a_{z}^{2}=18$, $m^{2}=3$.
 With $\kappa^{2}=5$,
 the axial function $Z_{a}(\varsigma)$ is shown in Fig.2,
 and a set of poloidal field lines is shown in Fig.3.

According to Eq.~\ref{eqno3},
 the poloidal and toroidal magnetic fields
 are given by
 
\begin{mathletters}
\begin{eqnarray}
\label{eqno25a}
\vec B_{p}\,
 =\,A_{0}{1\over r_{L}}{1\over\xi}\nabla\Psi\times\hat\phi
 \,\,\,,
\\
\label{eqno25b}
B_{\phi}\,
 =\,-\alpha A_{0}{k_{z}\over r_{L}}{1\over\xi}\Psi
 =\,-A_{\phi}{k_{z}\over r_{L}}{1\over\xi}\Psi
 \,\,\,,
\\
\label{eqno25c}
{\vec B_{p}\over B_{\phi}}\,
 =\,-{A_{0}\over A_{\phi}}{1\over k_{z}}{\nabla\Psi\over\Psi}
 \,\,\,,
\end{eqnarray}
\end{mathletters}

\noindent where the last equation
 gives the ratio of the two fields
 in the magnetosphere.
 However, this ratio is expressed
 in terms of their respective fluxes,
 $A_{0}$ and $A_{\phi}$,
 without which a specific ratio cannot be determined.

For the ion open field lines,
 we use the same representation of Eq.~\ref{eqno24},
 and keep $a_{\phi}^{2}=20$ and $\kappa^{2}=5$.
 However, we change $a_{L}^{2}=6$
 and other parameters become $a_{z}^{2}=16$, $m^{2}=1$.
 By changing $a_{L}^{2}=8$ to $a_{L}^{2}=6$,
 the radial function begins to present oscillations
 for small $\xi$,
 which is appropriate for the ion field lines. 
 The functions $R(\xi)$ and $Z_{a}(\varsigma)$
 are given in Fig.4 and Fig.5 respectively.
 We note that $R(\xi)$ rises from zero
 and passes through a maximum and a minimum locally
 before it takes off monotonically.
 The local maximum and minimum
 can become more pronounced
 by taking a smaller $a_{L}^{2}$.
 A set of poloidal field lines is shown in Fig.6
 which reflects the monopole-like solution
 of a charge separated magnetosphere in the near zone.
 In the boundary zone, the plasma tends to be quasi-neutral
 as it proceeds to meet the interstellar plasma in the far zone.
 The $\xi<1$ part of the open field lines
 is the solution of the pulsar equation,
 and the $\xi>1$ part is the analytic continuation across LC.
 
To construct the closed field lines of the ion plasma,
 known as the dead zone,
 we note that, with ion current outflow
 and electron return current outflow
 (closed at a distant load)
 along the open field lines equal in magnitude,
 $B_{\phi}=0$ in the closed region
 which requires $a_{z}^{2}=0$.
 To represent the closed field lines,
 we need a radial function
 bounded by a root $R(\xi)=0$ within LC.
 We also need a bounded axial function.
 We therefore write

\begin{eqnarray}
\label{eqno26}
\Psi(\xi,\varsigma)\,
 =\,R(\xi)Z'_{a}(\varsigma)\,
 =\,C\,\,\,.
\end{eqnarray}
 
\noindent In order to have $R(\xi)$
 bounded by a root within LC,
 we choose $a_{0}=+1$ and $a_{1}=-1.5$
 with different signs.
 Under the condition $a_{z}^{2}=0$,
 we take $a_{\phi}^{2}=-17$, $a_{L}^{2}=7$,
 and $\kappa'^{2}=5$ ($\kappa^{2}=-5$),
 we have $m^{2}=12$,
 and the functions $R(\xi)$ and $Z'_{a}(\varsigma)$
 are shown in Fig.7 and Fig.8 respectively.
 Once more, we recall that these parameters
 are neither unique ones nor restrictive.
 There is a good range in parameter space
 that reproduces the closed region. 
 The boundary condition of $B_{r}=0$
 on the equatorial plane at $\varsigma=0$
 is satisfied by $Z'_{a}(\varsigma)$.
 A corresponding set of poloidal field lines
 is shown in Fig.9
 with a different $\xi$ scale of Fig.3 and Fig.6
 for more clarity.
 In this figure, the closed region terminates
 at an intermediate value of $\xi=0.80$
 and the reason of which has been discussed in Sec.4.
 The same figure is redrawn with the same scale
 of Fig.3 and Fig.6,
 so that the three figures
 can be superimposed on each other
 to get the global features of the magnetosphere,
 as in Fig.10.
 This set of global field lines
 is a continuous function of the contour value $C$.
 More field lines could be added
 by assigning the intermediate contour values $C$.
 Comparing to the numerical results,
 such as \citet{contopoulos1999},
 our magnetospheric configuration
 reproduces well the main global features.

\newpage
\section{A Self-Consistent Description of Magnetosphere}

So far, we have introduced plasma drift velocities
 to remove the inconsistency
 of the space charge density
 with the closed field lines,
 and the incinsistency
 of the $E_{\parallel}=0$ condition
 with the emf driven plasma outflows.
 Furthermore, we have solved analytically
 the pulsar equation for the global structure
 of the magnetosphere.
 Above all, we have taken the view
 of a stationary magnetosphere in space
 under which the plasma drifts with a velocity $\vec v$
 according to Eq.~\ref{eqno2}.
 As we have already mentioned,
 the magnetosphere is described
 in terms of fields by Eq.~\ref{eqno1},
 and in terms of velocities by Eq.~\ref{eqno2}.
 We should note that the force-free approximation
 is only valid for the field-description.
 As for the velocity-description,
 plasma inertia and plasma pressure
 are the prime factors.
 These two descriptions are then connected
 by Eq.~\ref{eqno10}
 which describes the magnetosphere
 in terms of space charge densities.
 All these three aspects
 have to be self-consistent among each other.
 With a general poloidal and toroidal magnetic fields,
 the plasma velocities are given by Eq.~\ref{eqno8},
 derived from plasma drifts.
 As for the space charge density,
 it is reflected by the $\vec B\cdot\vec\Omega_{0}$
 on the right side of Eq.~\ref{eqno10}.
 
Before addressing the consistency of the space charge density,
 let us first establish the magnitudes
 of the $\vec v_{Rc}$ and $\vec v_{\nabla B}$ drifts
 by noting their ratio

\begin{eqnarray}
\label{eqno27}
{v_{Rc}\over v_{\nabla B}}\,
 =\,{L\over R_{c}}\,\,\,,
\end{eqnarray}
 
\noindent where $L$ is the scale length
 defined by $1/L=\nabla B/B$.
 Next, to compare with $\vec v_{E}$,
 we rewrite $\vec v_{E}$ in terms of $\Psi$
 which reads

\begin{eqnarray}
\label{eqno28}
\vec v_{E}\,
 =\,{1\over B^{2}}(\vec E_{\perp}\times\vec B)\,
 =\,-V_{0}{1\over B^{2}}(\nabla\Psi\times\vec B)\,\,\,.
\end{eqnarray}

\noindent For a poloidal magnetic field
 with $\Psi=C$ representing the field line,
 $\nabla\Psi$ is directed inward
 towards the equatorial plane,
 and $\vec v_{E}$ corresponds to
 a toroidal rotation of
 $\vec v_{E}=r\Omega_{0}\hat\phi$.
 For the $\vec v_{\nabla B}$ drift of Eq.~\ref{eqno9b},
 we can evaluate $\nabla B$
 by using the $\Psi=C$ representation to get
 
\begin{eqnarray}
\label{eqno29}
\nabla B\,=\,A_{0}a^{2}\nabla\Psi\,\,\,,
\end{eqnarray}

\noindent where $a$ has the dimension
 of an inverse scale length.
 Equivalently, $A_{0}a^{2}$ has the dimension
 of a reference poloidal mangnetic field.
 As a result, Eq.~\ref{eqno9b} reads
 
\begin{eqnarray}
\label{eqno30}
\vec v_{\nabla B}\,
 =\,-{\epsilon_{\perp}\over q}
 {A_{0}a^{2}\over B}{1\over B^{2}}
 (\nabla\Psi\times\vec B)\,\,\,.
\end{eqnarray}

\noindent We note that this equation is identical in form
 for a positive charge $q$ to Eq.~\ref{eqno28}.
 As for the $\vec v_{Rc}$ drift of Eq.~\ref{eqno9a},
 we note that the radius of curvature $\vec R_{c}$
 directs from the center of curvature
 outward to the field lines.
 Therefore, $\vec R_{c}$ and $-\nabla\Psi$
 are pointing outward,
 although they are not parallel.
 Consequently, both $\vec v_{\nabla B}$ and $\vec v_{Rc}$
 drifts are in the toroidal direction,
 as is $\vec v_{E}$.

As for their amplitudes,
 based on Eq.~\ref{eqno28}, Eq.~\ref{eqno30}, and Eq.~\ref{eqno9a},
 all three velocities have the same $1/B^{2}$ scaling,
 so their strengths can be compared by other factors.
 For $\vec v_{Rc}$, we note that $R_{c}$ is small
 when it is near the pulsar
 and when it is near the apex
 on returning back to the conjugate point
 of the southern equatorial region.
 In between, the field lines are relatively flat,
 as shown in Fig.9, and $R_{c}$ is large.
 As for $\vec v_{\nabla B}$,
 the ratio of this velocity to $\vec v_{E}$ is
 
\begin{eqnarray}
\label{eqno31}
{v_{E}\over v_{\nabla B}}\,
 =\,{qV_{0}\over\epsilon_{\perp}}
 {B\over A_{0}a^{2}}\,\,\,.
\end{eqnarray}

\noindent Assumming the reference poloidal field $A_{0}a^{2}$
 comparable to $B$,
 these two velocities become comparable
 when the plasma energy (pressure) $\epsilon_{\perp}=qV_{0}$.
 The same can also be said
 for $\epsilon_{\parallel}$ for $\vec v_{Rc}$.
 This high plasma pressure can be found
 close to the pulsar surface,
 and it decreases as $r$ increases.

For a toroidal magnetic field in $+\hat\phi$ direction,
 the $\vec v_{E}$ drift will be poloidal
 but in the counter direction of closed field lines.
 Likewise are $\vec v_{\nabla B}$ and $\vec v_{Rc}$.
 By Eq.~\ref{eqno6}, we therefore have $\kappa_{v}<0$,
 and the factor $(\vec v_{E}-\kappa_{v}\vec B)$
 in Eq.~\ref{eqno10} is positive.
 Should we consider the toroidal magnetic field
 be in $-\hat\phi$ direction,
 we would get $\kappa_{v}>0$,
 and the factor $(\vec v_{E}-\kappa_{v}\vec B)$
 would be positive again.
 We remark that the space charge density of Eq.~\ref{eqno10}
 is valid for the entire magnetosphere,
 not just for the closed region.
 In the presence of poloidal and toroidal fields,
 the scalar product in the square bracket
 on the right side of Eq.~\ref{eqno10}
 has a poloidal and a toroidal conribution.
 This space charge density has to be calculated
 together with the fields and plasma drifts
 in numerical solutions.

\newpage
\section{Pulsating Magnetosphere and Magnetosphere Instability}

We have constructed the standard magnetsphere in Sec.7
 where the closed region with $B_{\phi}=0$
 is described by Eq.~\ref{eqno26}.
 On the other hand,
 with the self-consistent description
 of the magnetosphere
 based on fields, velocities,
 and space charge densities in Sec.8,
 there should be a $B_{\phi}$ in the closed region.
 On this question of a toroidal field,
 we note that, in a charge separated magnetosphere,
 a toroidal magnetic field can arise spontaneously.
 If we begin with a pure poloidal magnetic field
 in the polar region open electron field lines,
 equilibrium condition requires
 the electron plasma be at rest.
 This is an unstable configuration,
 since any longitudinal movement of the electron plasma
 corresponds to a poloidal current
 that would generate a toroidal magnetic field.
 This toroidal magnetic field would enhance
 the poloidal drift velocity (current) itself.
 Since the field lines are open,
 an equilibrium configuration
 of fields and currents could be reached.
 With $\rho_{q}<0$ for electron field lines,
 and with small $\vec v_{\phi}$ due to small $\xi$,
 it is reasonable that the square bracket
 be positive and thus $B_{z}>0$.
 As for the ion open field lines,
 the toroidal field here begins to decline
 due to the ion outflows.
 With $\rho_{q}>0$ for ion field lines,
 it is likely that we could have
 $B_{z}>0$ in some part of the field line and
 $B_{z}<0$ in other part.

As for the ion field lines in the closed region,
 there still is a toroidal field
 due to the imperfect shielding
 of the electron axial current
 by the ion outflow in the open region.
 With $\rho_{q}>0$, and with the presence of a toroidal field
 and a high plasma pressure near the pulsar,
 Eq.~\ref{eqno10} could have $B_{z}>0$ near the pulsar.
 Farther out with large $\xi$,
 the toroidal field becomes small,
 likewise is the plasma pressure,
 we could expect $B_{z}<0$.
 With the poloidal field lines described by Eq.~\ref{eqno26},
 the corresponding toroidal field in the closed region
 is represented by Eq.~\ref{eqno25b}.
 Because of the mirror symmetry
 between the upper (north) and lower (south) hemispheres,
 the toroidal fields have opposite sign in the two parts,
 which drives an outward radial current sheet
 on the equatorial plane.
 On reaching the apex of the closed region,
 this current sheet divides into two parts
 returning from the northern and southern
 closed poloidal field lines.
 We therefore have two current loops,
 one north and one south,
 to enhance the corresponding toroidal fields.
 This in turn enhances the radial current sheet
 and the current loops.
 In a closed finite region,
 this positive feedback cycle can only be stablized
 in the presence of dissipations.
 Naturally, there is the usual toroidal current sheet
 on the equatorial plane beyond LC
 for the split monopole poloidal field lines.

To find an alternative beacon mechanism,
 we first racall that the current model
 pictures a corotating magnetic field
 that drags the magnetospheric plasma.
 In response to this,
 a rotation induced electric field
 is generated in the stationary frame of space.
 In this picture,
 the rotating magnetic field is the prime driver,
 and the magnetosphere is decoupled
 from the unipolar induction of the pulsar.
 Here, we view the unipolar induction inside the pulsar
 and the magnetosphere outside the pulsar
 as one coupled system.
 The unipolar induction delivers charges
 separated by their signs
 to the polar and equatorial regions
 at a given rate.
 The current flow inside the neutron star
 interacts with the magnetic field,
 generating a braking torque.
 The charges are then channelled to the magnetosphere
 along the field lines,
 and are ejected to the interstellar space
 as stellar winds through open field lines,
 which allows the unipolar induction to operate continuously.
 Considering the unipolar energy pumping rate
 exceeds the Poynting outflow rate
 along the open field lines,
 plasma density and pressure will build up
 in the magnetosphere,
 particularly in the closed region.
 In the outer part of the closed region
 extending to intermediate $\xi$ with $B_{z}<0$,
 $\vec v_{\nabla B}$ and $\vec v_{Rc}$ are small.
 However, as the plasma pressure build-up reaches there,
 these velocities begin to increase.
 Eventually, they get to the point of having $B_{z}>0$,
 flipping open the field lines
 and releasing the trapped plasma and magnetic energies.
 Furthermore, the field lines can also be flipped open
 in the outer part of the closed region
 by the positive feedback on the toroidal field
 which appears in the $(\vec v_{E}-\kappa_{v}\vec B)$ factor
 of Eq.~\ref{eqno10}.

Once depleted of plasmas,
 in the sense that plasma density is so low
 that Eq.~\ref{eqno2} is not warranted,
 the magnetosphere will then be recharged all over
 to the standard configuration,
 and the whole cycle starts anew.
 Instead of a stationary magnetosphere,
 we therefore have a scenario
 of a pulsating magnetosphere
 where the openning of the closed field lines
 either by unipolar pumping on drift velocities
 or by positive feedback on the toroidal field
 works as a magnetic switch,
 which offers an altenative mechanism for pulsar beacons.
 In this alternative picture,
 an observer aligned with the magnetic axis
 sees the beacons separated
 by the unipolar induction charging period,
 instead of the pulsar rotation period.

Here, we have reached the pulsating picture
 of the magnetosphere
 through a steady state approach
 by viewing the unipolar induction inside the pulsar
 and the magnetosphere outside as a coupled system.
 As a matter of fact,
 the steady state approach
 may not be adequate to describe
 a charge separated magnetosphere.
 We call to the attention
 that a charge separated plasma
 is a very unstable plasma due to the fact
 that any plasma flow (drift) is a current flow
 which generates magnetic field that drives drifts.
 For this reason,
 we might have to let go the lighthouse paradigm,
 and consider the time dependent model
 with Maxwell equations coupled to the plasma equations
 \citep{komissarov2006, spitkovsky2006, tchekhovskoy2013, urpin2012, urpin2014}.
 Furthermore, the inherent unstable nature
 of a charge separated magnetosphere
 could be the cause of periodic beaming
 of electromagnetic waves.

\newpage
\section{Conclusions}

We have taken the view of interpreting
 the magnetic field $\vec B$
 of the magnetosphere as stationary in space,
 and the electric field $\vec E$
 as established by uploading
 the potential distribution of the pulsar surface
 to the magnetic field lines.
 The plasma velocity $\vec v$
 is, therefore, the drift response in Eq.~\ref{eqno2}
 which rotates with an angular velocity of the pulsar.
 The presence of a magnetosphere
 allows the pulsar surface charges
 be channelled to the field lines,
 and keeps the counter electric field
 in the pulsar interior low.
 This warrants the continuous operation
 of the unipolar induction
 that constantly pumps energy to the magnetosphere.
 In this scenario,
 the energetics of the magnetosphere
 is coupled to the unipolar induction
 of the neutron star.

By introducing the gradient B
 and curvature plasma drifts
 and the presence of a toroidal field,
 we have removed the inconsistency
 between the space charge density
 and the curvature of the dipole-like
 closed ion field lines.
 Also, the inconsistency
 between introducing the far zone
 interstellar plasma floating potential
 to the near zone open field lines
 to generate an emf driven
 plasma flow along open field lines
 and the $E_{\parallel}=0$ plasma condition
 can be removed as well by drift velocities.
 Although plasma inertia and pressure are neglected
 in the force-free pulsar equation,
 they enter the magnetospheric dynamics implicitly
 through the $\vec E\times\vec B$, gradient B,
 and curvature plasma drifts,
 which are the respective guiding center drifts
 of the basic gyro motion.
 Since the gradient B and curvature drifts are mass dependent,
 the closed region should be bounded
 at a distance within LC.

We have devised a method to solve
 the pulsar equation analytically
 for the standard split monopole magnetosphere.
 The solutions are given in terms of three parameters,
 the separation function parameter $f(\Psi)=k^{2}\Psi$,
 the axial current parameter $A(\Psi)=-k_{z}\Psi$,
 and the separation constant $m^{2}$.

Recognizing that the magnetosphere
 is described by the fields of Eq.~\ref{eqno1}
 under force-free approximation,
 and by the plasma drift velocities of Eq.~\ref{eqno2}
 where plasma inertia and plasma pressure
 are prime factors,
 and these two descriptions are connected
 by the space charge densities of Eq.~\ref{eqno10},
 we have presented a self-consistent description
 of the pulsar magnetosphere.
 Through the coupled system
 between the unipolar induction inside the neutron star
 and the magnetosphere outside,
 and considering the unipolar induction rate
 exceeds the open field line Poynting flux,
 we have constructed a dynamic pulsating magnetosphere.
 In this pulsating model,
 the accumulated magnetic and plasma energies
 of the closed field lines
 can be released periodically
 by switching open the field lines
 either as plasma pressure gets high enough
 or as the toroidal magnetic field gets large enough.
 As an alternative to the lighthouse paradigm,
 this pulsating magnetosphere
 offers a different mechanism
 to generate pulsar beacons.

\acknowledgments

\clearpage
\begin{figure}
\plotone{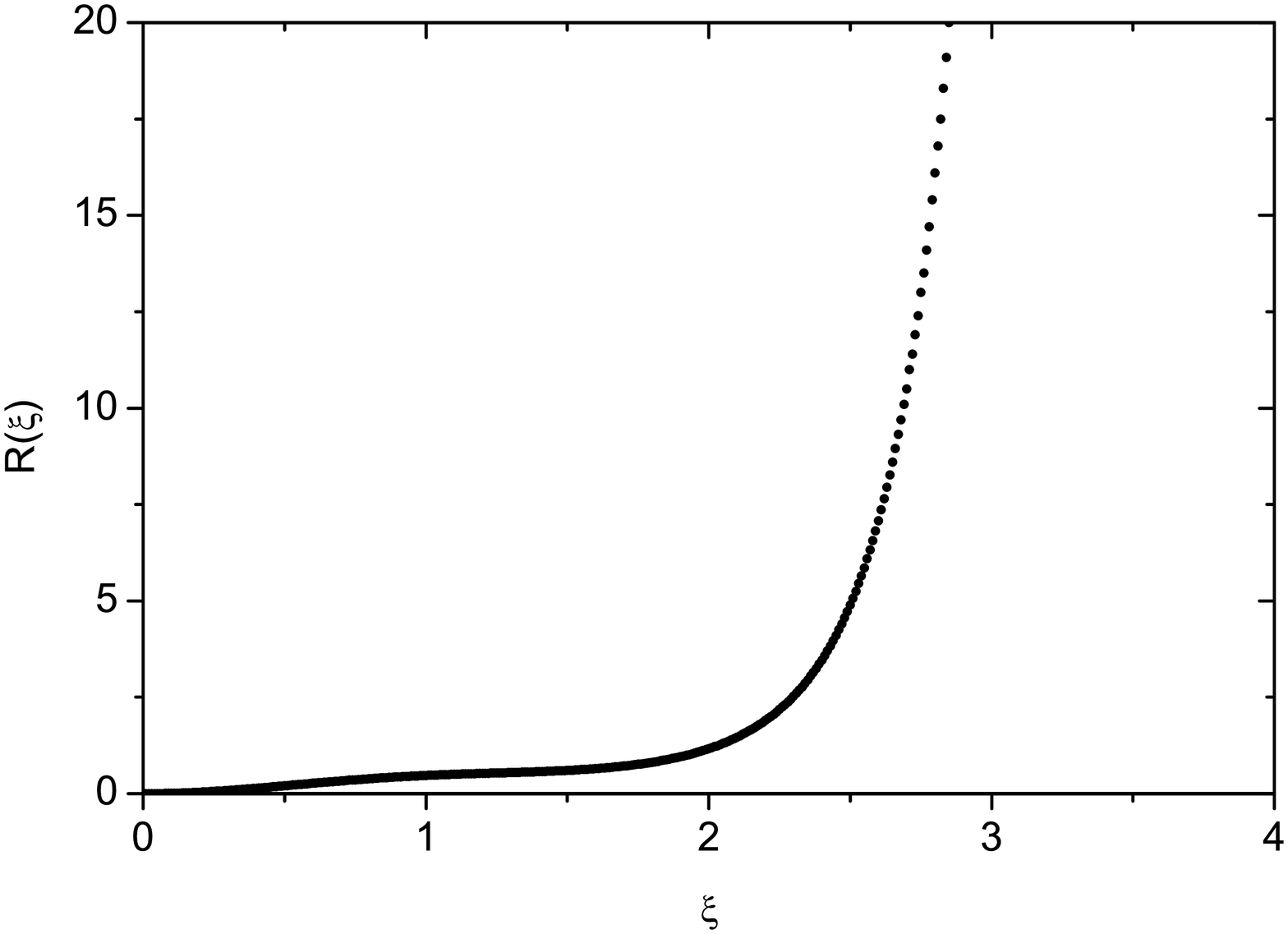}
\caption{With $a_{0}=1$ and $a_{1}=0$,
 and $a_{\phi}^{2}=20$ and $a_{L}^{2}=8$,
 the radial solution $R(\xi)$ with monotonically increasing profile
 is plotted as a function of the normalized radial distance $\xi$.}
\label{fig.1}
\end{figure}

\clearpage
\begin{figure}
\plotone{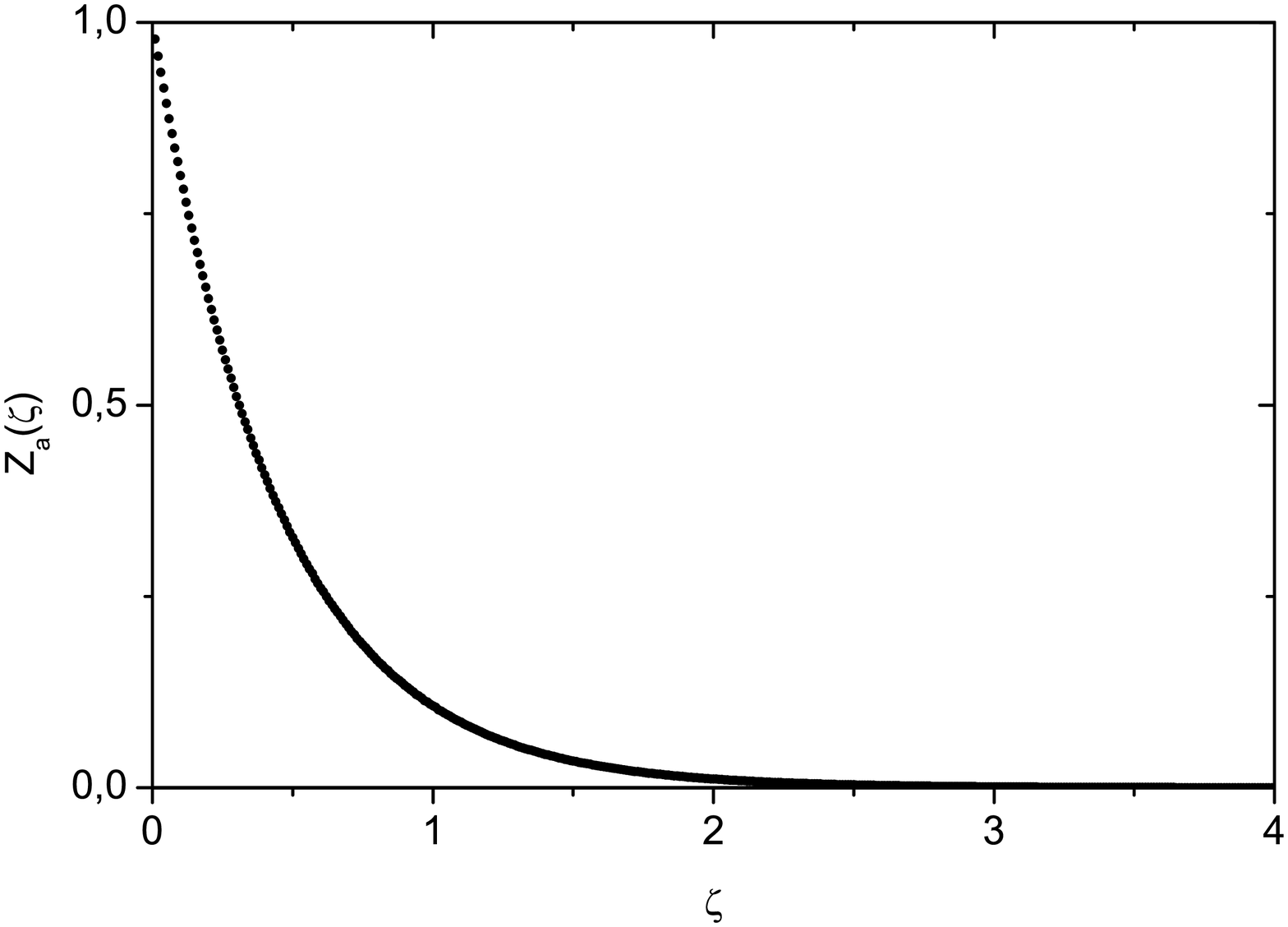}
\caption{With $\kappa^{2}=5$
 corresponding to $a_{L}^{2}=8$ and $m^{2}=3$,
 the axial function $Z_{a}(\varsigma)$ with exponentially decaying profile
 is plotted as a function of the normalized axial distance $\varsigma$.}
\label{fig.2}
\end{figure}

\clearpage
\begin{figure}
\plotone{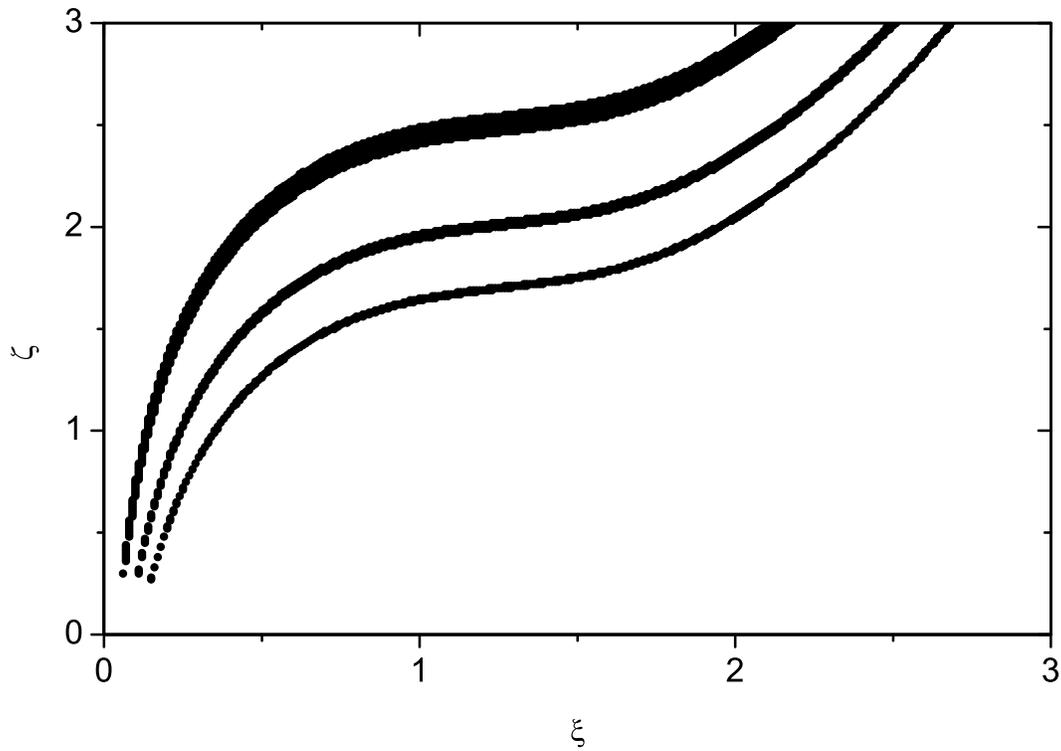}
\caption{A set of poloidal flux function $\Psi(\xi,\varsigma)$
 for the polar region electron open field lines
 is illustrated with contour values $C=0.002,\,0.006,\,0.012$
 ordering from the polar axis towards the equator.
 The radius of pulsar is taken as $\xi_{pulsar}=0.3$.}
\label{fig.3}
\end{figure}

\clearpage
\begin{figure}
\plotone{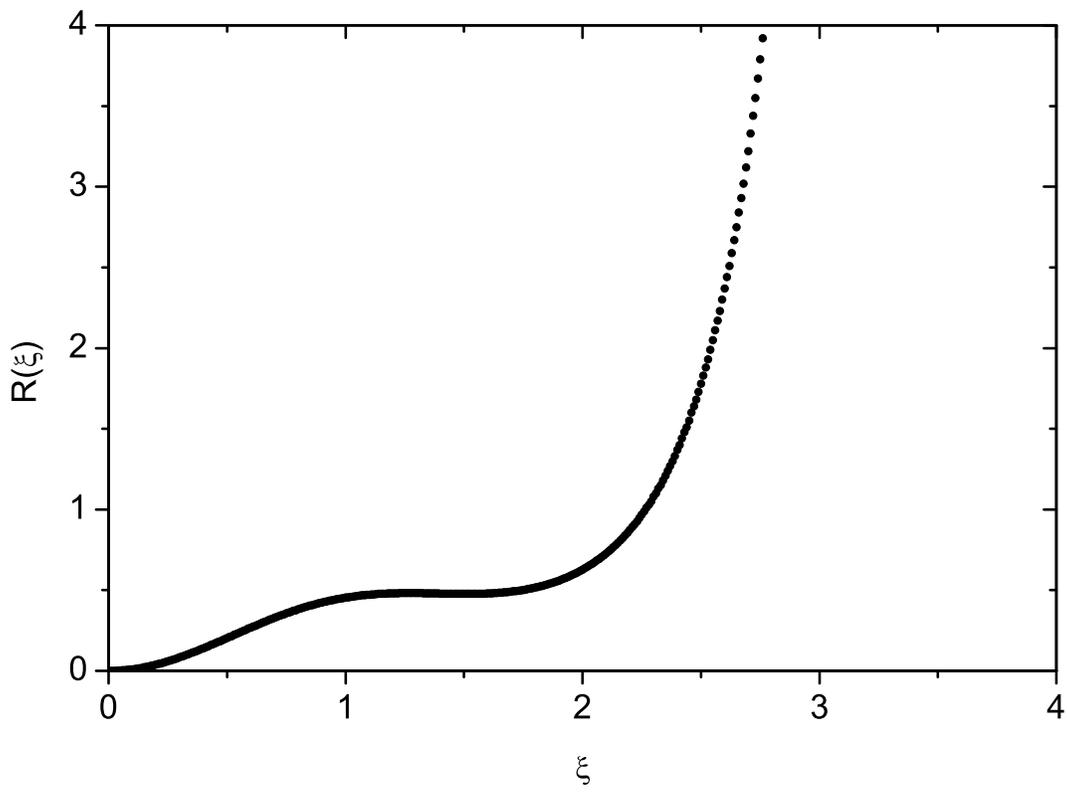}
\caption{With $a_{0}=1$ and $a_{1}=0$,
 and $a_{\phi}^{2}=20$ and $a_{L}^{2}=6$,
 the radial solution $R(\xi)$ with a local maximum and minimum
 is plotted as a function of the normalized radial distance $\xi$.}
\label{fig.4}
\end{figure}

\clearpage
\begin{figure}
\plotone{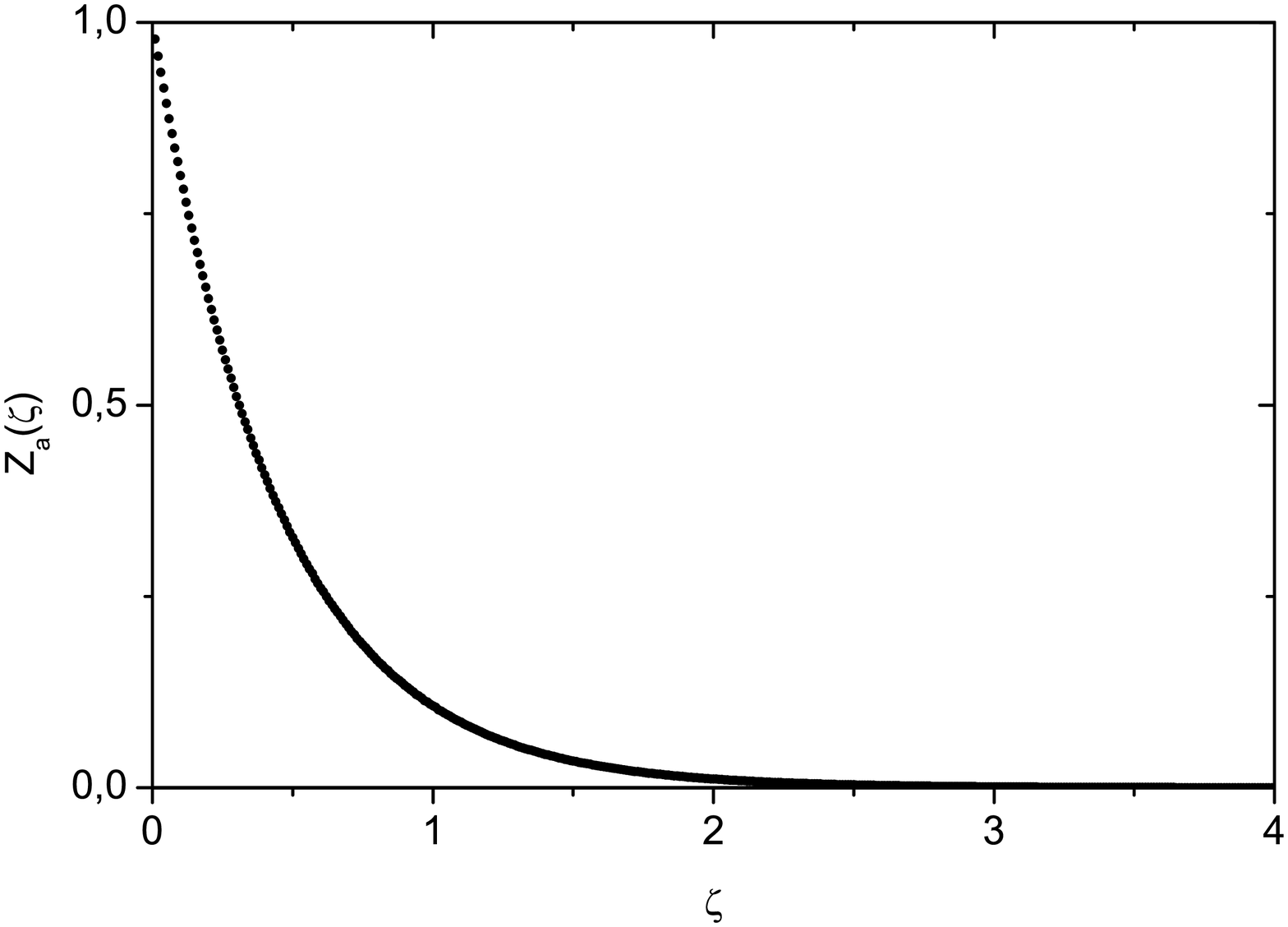}
\caption{With $\kappa^{2}=5$
 corresponding to $a_{L}^{2}=6$ and $m^{2}=1$,
 the axial function $Z_{a}(\varsigma)$ with exponentially decaying profile
 is plotted as a function of the normalized axial distance $\varsigma$.}
\label{fig.5}
\end{figure}

\clearpage
\begin{figure}
\plotone{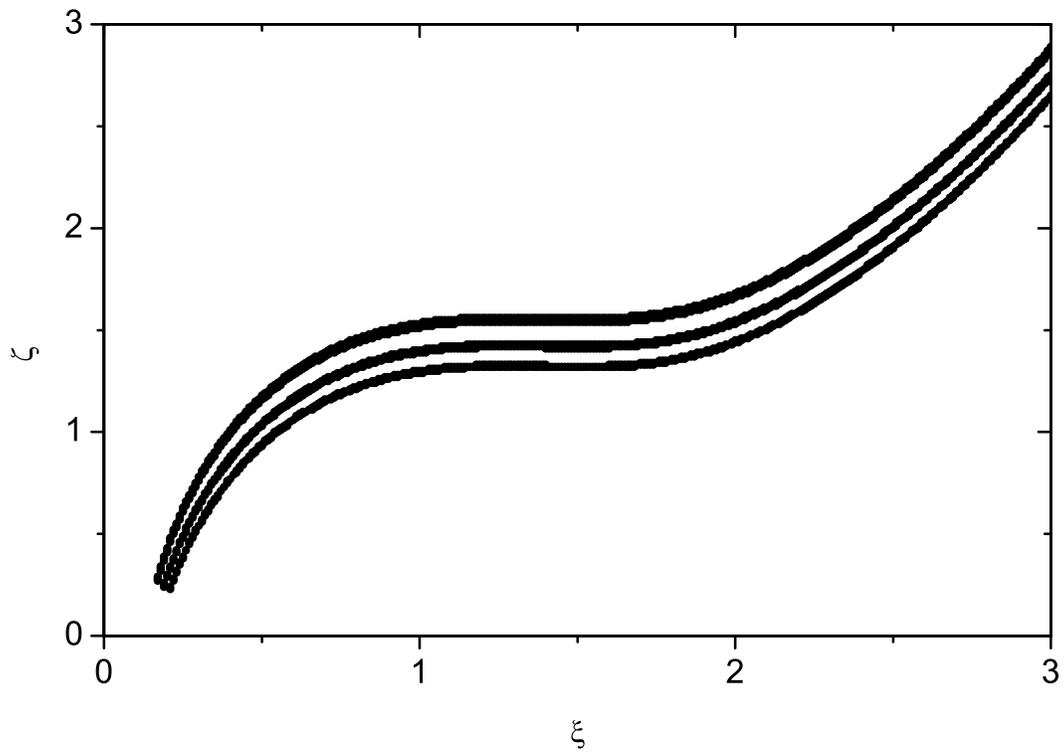}
\caption{A set of poloidal flux function $\Psi(\xi,\varsigma)$
 for the equatorial region ion open field lines
 is illustrated with contour values $C=0.015,\,0.020,\,0.025$
 with the same ordering.
 These field lines become monopole-like for large $\xi$.}
\label{fig.6}
\end{figure}

\clearpage
\begin{figure}
\plotone{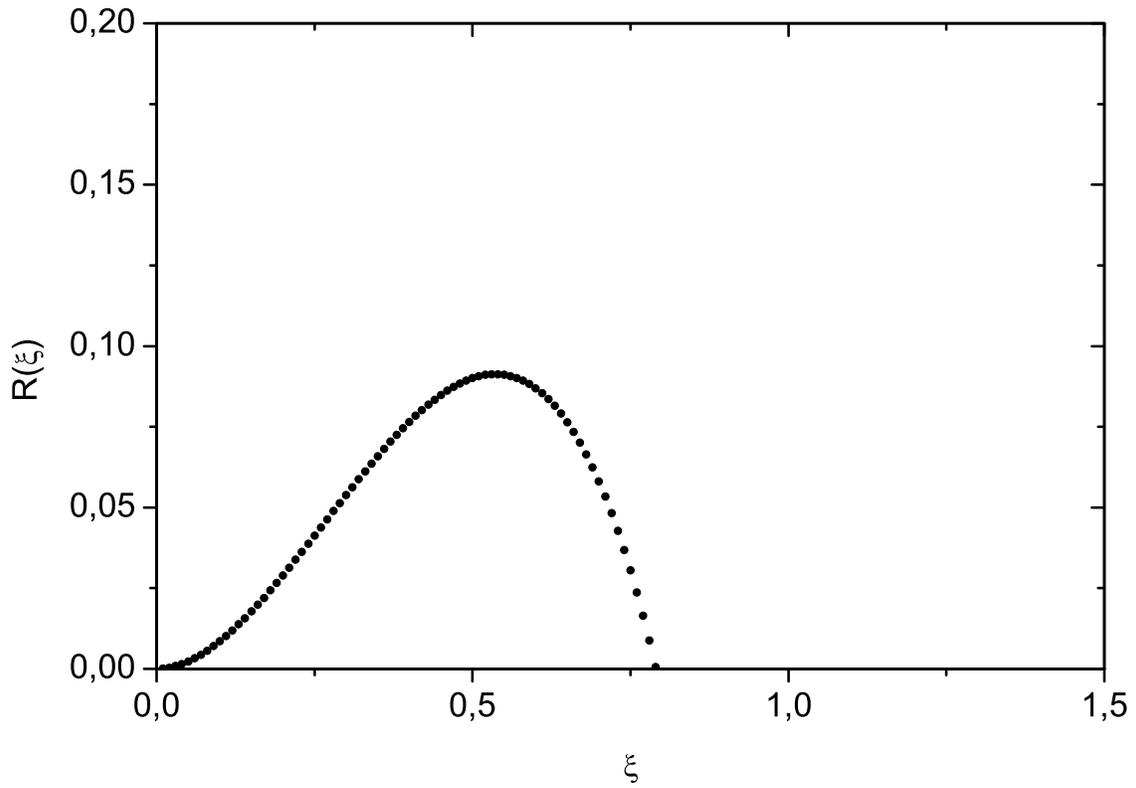}
\caption{With $a_{0}=1$ and $a_{1}=-1.5$,
 and $a_{\phi}^{2}=-17$ and $a_{L}^{2}=7$,
 plus the condition of $a_{z}^{2}=0$,
 the radial solution $R(\xi)$ with a root within LC
 is plotted as a function of the normalized radial distance $\xi$.}
\label{fig.7}
\end{figure}

\clearpage
\begin{figure}
\plotone{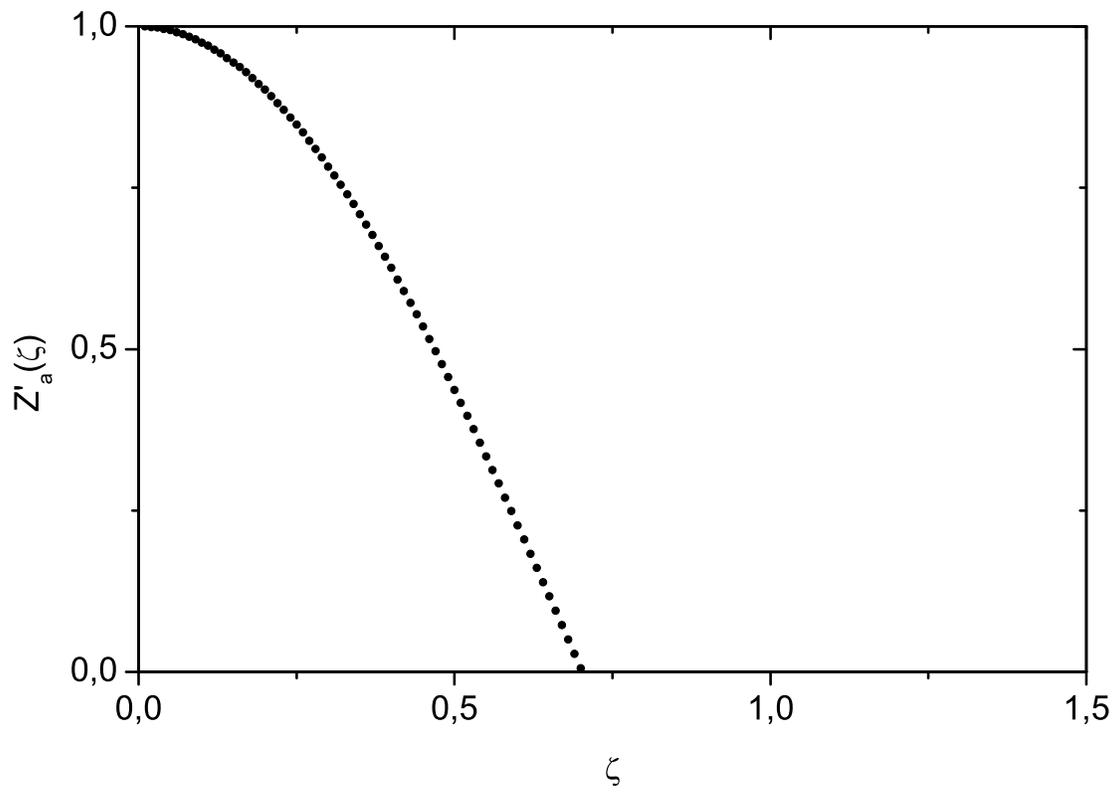}
\caption{With $\kappa'^{2}=5$ ($\kappa^{2}=-5$)
 corresponding to $a_{L}^{2}=7$ and $m^{2}=12$,
 the axial function $Z'_{a}(\varsigma)$ with cosine profile
 is plotted as a function of the normalized axial distance $\varsigma$.}
\label{fig.8}
\end{figure}

\clearpage
\begin{figure}
\plotone{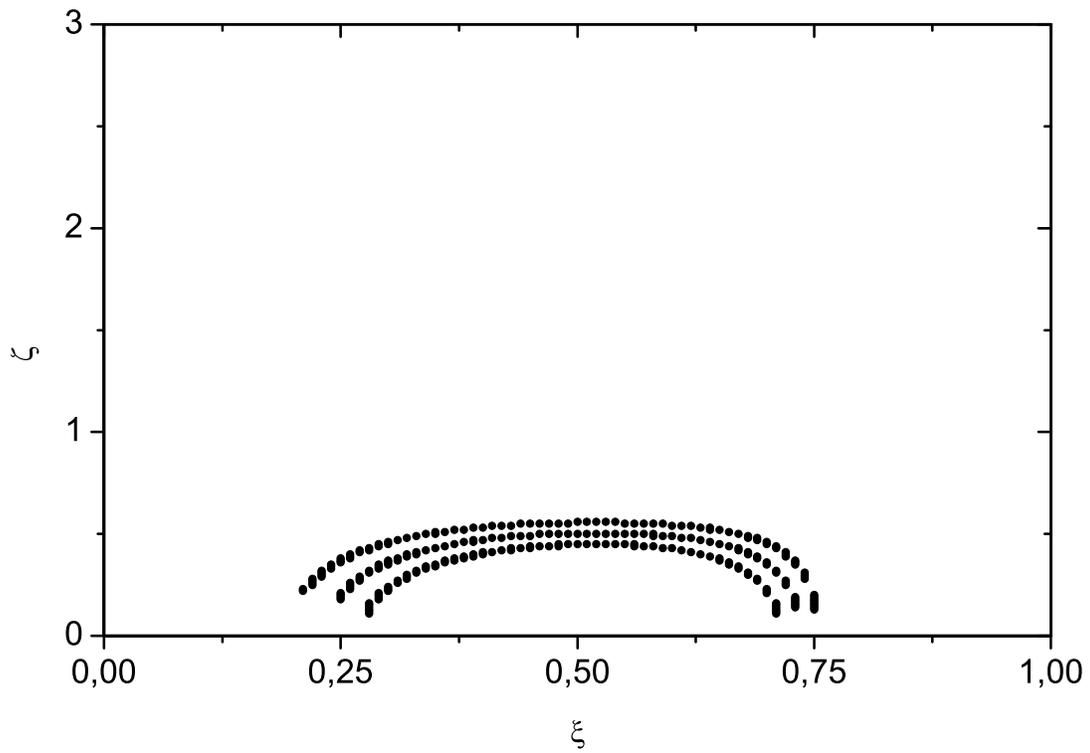}
\caption{A set of poloidal flux function $\Psi(\xi,\varsigma)$
 for the equatorial region ion closed field lines
 is illustrated with contour values $C=0.03,\,0.04,\,0.05$
 with the same ordering.
 These field lines extend to an intermediate range of $\xi$
 to assure $B_{z}<0$ as required by the space charge density.}
\label{fig.9}
\end{figure}

\clearpage
\begin{figure}
\plotone{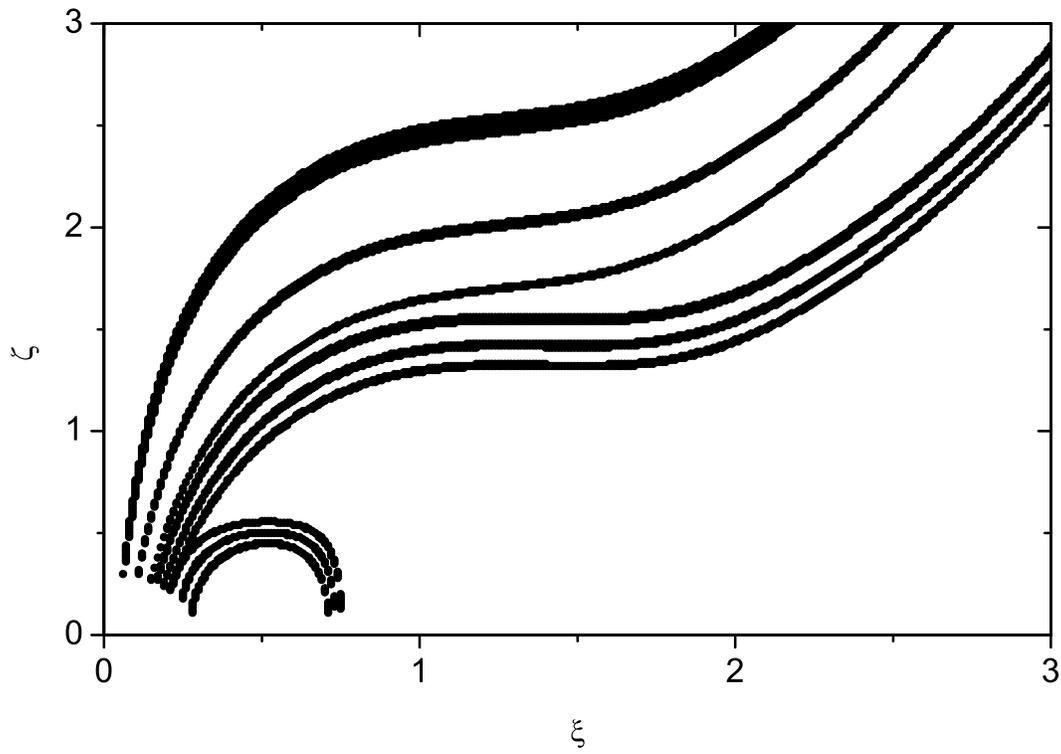}
\caption{The electron and ion open field lines
 and the ion closed field lines
 of Figs.3,6,9 are superpositioned
 to show the global structure of the pulsar magnetosphere.
 This global set of field lines
 is a continuous function of the contour valus $C$.}
\label{fig.10}
\end{figure}


\newpage
\begin{thebibliography}{}

\bibitem[Beskin \& Malyshkin(1998)]{beskin1998}
 Beskin, V.S. \& Malyshkin, L.M., 1998.
 {On the self-consistent model of an axisymmetric
  radio pulsar magnetosphere},
 \mnras, {\bf{298}}, 847-853.

\bibitem[Contopoulos(2005)]{contopoulos2005}
 Contopoulos, I., 2005.
 {The coughing pulsar magnetosphere},
 \aap, {\bf{442}}, 579-586.

\bibitem[Contopoulos et al.(1999)]{contopoulos1999}
 Contopoulos, I., Kazanas, D., \& Fendt, C., 1999.
 {The axisymmetric pulsar magnetosphere},
 \apj, {\bf{511}}, 351-358.

\bibitem[Fitzpatrick \& Mestel(1988)]{fitzpatrick1988}
 Fitzpatrick, R. \& Mestel, L., 1988.
 {Pulsar electrodynamics - I},
 \mnras, {\bf{232}}, 277-302.

\bibitem[Gold(1968)]{gold1968}
 Gold, T., 1968.
 {Rotating Neutron Stars as the Origin of the Pulsating Radio Sources},
 Nature, {\bf{218}}, 731-732.

\bibitem[Goldreich \& Julian(1969)]{goldreich1969}
 Goldreich, P. \& Julian, W.H., 1969.
 {Pulsar Electrodynamics},
 \apj, {\bf{157}}, 869-880.

\bibitem[Goodwin et al.(2004)]{goodwin2004}
 Goodwin, S.P., Mestel, J., Mestel, L., \& Wright, G.A.E., 2004.
 {An idealized pulsar magnetosphere :
  the relativistic force-free approximation},
 \mnras, {\bf{349}}, 213-224.

\bibitem[Gruzinov(2005)]{gruzinov2005}
 Gruzinov, A., 2005.
 {Power of an axisymmetric pulsar},
 \prl, {\bf{94}}, 021101.

\bibitem[Jackson(1975)]{jackson1975}
 Jackson, J.D., 1975.
 {Classical electrodynamics},
 Chapter 12, John Wiley \& Sons, New York.

\bibitem[Jackson(1976)]{jackson1976}
 Jackson, E.A., 1976.
 {A new pulsar magnetospheric model.
 I. Aligned magnetic and rotational axes},
 \apj, {\bf{206}}, 831-841.

\bibitem[Komissarov(2006)]{komissarov2006}
 Komissarov, S.S., 2006.
 {Simulations of the axisymmetric magnetospheres of neutron stars},
 \mnras, {\bf{367}}, 19-31.

\bibitem[Melrose(1978)]{melrose1978}
 Melrose, D.B., 1978.
 {Amplified liner acceleration emission applied to pulsars},
 \apj, {\bf{225}}, 557-573.

\bibitem[Mestel et al.(1985)]{mestel1985}
 Mestel, L., Robertson, J.A., Wang, Y.M., \& Westfold, K.C., 1985.
 {The axisymmetric pulsar magnetosphere},
 \mnras, {\bf{217}}, 443-484.

\bibitem[Michel(1979)]{michel1979}
 Michel, F.C., 1979.
 {Vacuum gaps in pulsar magnetospheres},
 \apj, {\bf{227}}, 579-589.

\bibitem[Ogura \& Kojima(2003)]{ogura2003}
 Ogura, J. \& Kojima, Y., 2003.
 {Some properties of an axisymmetric pulsar magnetosphere
  constructed by numerical claculations},
 Prog. Theo. Phys., {\bf{109}}, 619-630.

\bibitem[Okamoto(1974)]{okamoto1974}
 Okamoto, I., 1974. 
 {Force-free pulsar magnetosphere - 
 I. The steady axisymmetric theory for the charge-separated plasma},
 \mnras, {\bf{167}}, 457-474.

\bibitem[Ruderman \& Sutherland(1975)]{ruderman1975}
 Ruderman, M.A. \& Sutherland, P.G., 1975.
 {Theory of Pulsars : 
 Polar gaps, sparks,and coherent microwave radiation},
 \apj, {\bf{196}}, 51-72.

\bibitem[Scharlemann \& Wagoner(1973)]{scharlemann1973}
 Scharlemann, E.T. \& Wagoner, R.V., 1973.
 {Aligned rotating magnetospheres. I. General analysis},
 \apj, {\bf{182}}, 951-960.

\bibitem[Schmidt(1966)]{schmidt1966}
 Schmidt, G., 1966.
 {Physics of high temperature plasmas},
 Chapter II, Academic Press, New York.

\bibitem[Spitkovsky(2006)]{spitkovsky2006}
 Spitkovsky, A., 2006.
 {Time-dependent force-free pulsar magnetospheres :
 axisymmetric and oblique rotators},
 \apj, {\bf{648}}, L51-L54.

\bibitem[Sturrock(1971)]{sturrock1971}
 Sturrock, P.A., 1971.
 {A model of pulsars},
 \apj, {\bf{164}}, 529-556.

\bibitem[Tchekhovskoy et al.(2013)]{tchekhovskoy2013}
 Tchekhovskoy, A., Spitkovsky, A., \& Jason, G.L., 2013.
 {Time-dependent 3D magnetohydrodynamic pulsar magnetospheres :
 oblique rotators},
 \mnras, {\bf{435}}, L1-L5.

\bibitem[Thompson(1962)]{thompson1962}
 Thompson, W.B., 1962.
 {An introduction to plasma physics},
 Chapter 7, Addison-Wesley, London.

\bibitem[Timokhin(2006)]{timokhin2006}
 Timokhin, A.N., 2006.
 {On the force-free magnetosphere of an aligned rotator},
 \mnras, {\bf{368}}, 1055-1072.

\bibitem[Urpin(2012)]{urpin2012}
 Urpin, V., 2012.
 {Force-free pulsar magnetosphere :
 instability and generation of magnetohydrodynamic waves},
 \aap, {\bf{541}}, A117.

\bibitem[Urpin(2014)]{urpin2014}
 Urpin, V., 2014.
 {Formation of filament-like structures in the pulsar magnetosphere
 and the short-term variability of pulsar emission},
 \aap, {\bf{563}}, A29.

\end{thebibliography}
\end{document}